\documentclass[preprint2]{aastex}

\begin{document}

\title{Upper Limits on the X-ray Emission of ``Uranium'' Stars}

\author{Eric M. Schlegel}

\affil{Harvard-Smithsonian Center for Astrophysics}


\begin{abstract} 

A paper by Qian \& Wasserburg suggests the optical absorption lines of
uranium observed in the spectra of ultra-metal-poor stars (defined as
[Fe/H] $<$-3) arise from contamination from a supernova in a binary
star system.  Assuming the binary survived the explosion, a collapsed
compact object may be present and implying potential accretion
processes with accompanying X-ray emission.  Upper limits on X-ray
emission from an accreting compact object are described here.

\end{abstract}


\section{Introduction}

Datings of radioactive materials provide the firmest ages known.  In
the case of long-lived heavy elements such as thorium and uranium,
derived age values encompass the estimated age of the Galaxy.

Ultra-metal poor stars, defined as stars with very low abundances of
heavy elements ([Fe/H] $\sim$-3) and largely confined to the Galactic
halo, evidently represent the earliest generation of stars formed as
the Galactic disk was still collapsing (e.g.,
\citealt{QW2001a,Cay01}).  As such, they are largely devoid of
contamination by subsequent generations of supernovae and star
formation that occurred within the disk.

Recently, observations of three stars, uncovered from optical surveys
of ultra-metal-poor stars (\citealt{NRB96} and references therein),
revealed absorption lines of r-process elements.  BPS CS31082-001,
hereafter CS31082, exhibits an optical absorption line from uranium at
3889.57{\AA} which yields an abundance of log(U/H) =
-13.7$\pm$0.14$\pm$0.12 (random + systematic) and an estimated age for
the star of 12.5$\pm$3 Gyr \citep{Cay01}.  BPS CS22892-052, hereafter
CS22892, also exhibits optical absorption lines from heavy elements,
particularly thorium (absorption line at 4019{\AA}, \citealt{Sned00}).
Both stars show absorption lines from r-process elements enhanced
relative to Fe.  \cite{Cow02} report the detection of a weak optical
line in BD+17$^{\circ}$ 3248, hereafter BD17, at 3589.6{\AA} from U
II.  Other stars undoubtedly show similar line features of r-process
elements.

Uranium and other neutron-rich nuclei have long been believed to be
produced in the explosive nucleosynthesis accompanying a supernova
(e.g., \citealt{HNW84}).  Briefly, the r-process is a chain of neutron
captures required to explain neutron-rich nuclides for species above
atomic mass 70. Successful production requires very high neutron
densities (N$_{\rm n}$ $>$10$^{19}$ cm$^{-3}$) and high temperature (T
$\sim$10$^9$ K) for the capture process to proceed more rapidly than
${\beta}$ decay.  The site of the r-process is not well-defined, but
the required conditions essentially demand an association with a
supernova.

Subsequent research raised significant questions about the prompt
mechanism and spurred searches for other sites.  One mechanism
currently enjoying wide support sites the r-process in the
$\nu$-heated ejecta of the newborn neutron star (\citealt{WH92, Mey92,
Woos94, Otsuki00, Tera01}).  Other sites include merging neutron stars
\citep{FRT99} and collapsing ONeMg cores \citep{WCH98}.  \cite{Sumi01}
resurrected the prompt mechanism using new, slower e$^-$-capture
rates.  A recent paper by \cite{Mey02} shows that an excess of
neutrons may not be necessary for the r process to proceed.

\cite{QW2001b} argued that the presence of r-processed material in
ultra-metal-poor stars suggested the possible presence of a compact
binary companion that formed after the detonation of the supernova.
Their argument is based upon a theory for element production developed
during the past few years (summarized in \citealt{QW2000}).  Based
upon their study of the cosmic abundances, they postulate the
existence of high- and low-frequency Type II (core collapse)
supernovae.  The class of progenitors of the high-frequency supernovae
detonate quickly, on the time scale of stellar evolution, or with a
rate of approximately 10$^{-7}$ yr$^{-1}$.  The low-frequency
supernovae detonate less often, at a rate of approximately 10$^{-8}$
yr$^{-1}$.

The nucleosynthetic products generated by the two types of supernovae
differ, with the frequent supernovae producing the majority of the
heavy r-process nuclei (with mass number $>$ 135) while the
less-frequent supernovae account for the majority of the lighter
r-process nuclei (mass number $<$135).  Evidence in support for the
bifurcation in r-process production was presented by \cite{Sned00} who
showed that the abundances of CS22892-052 followed the solar r-pattern
for elements heavier than barium, but were difficient for elements
near Ag (A = 107).

\cite{WQ2000} argue that the presence of heavy r-process elements in
ultra-metal-poor stars with low Fe abundances implies that the
high-frequency supernovae can produce r-process material {\it without}
the existence of a large abundance of iron.  The observations of
uranium in CS31082-001 provide evidence in support for their argument
\citep{QW2001b}.

If a compact companion did form and if the binary survived the
explosion, the companion's presence should be detected from the
production of X-rays from accretion processes.  Sensitive X-ray
observations may test for this situation.

\section{Observations}

We surveyed the entire data base of satellite X-ray data available
through the HEASARC collection.  No pointed observations from
Einstein, ROSAT, ASCA, Chandra, or XMM-Newton exist within 60$'$ of
the targets.  The ROSAT All-Sky Survey\footnote{We acknowledge with
thanks the support for the RASS provided by Max-Planck Institut f\"ur
Extraterrestriche Astrophysik.  The research discussed here would have
been impossible without that support.} (RASS) provides the only
measurements of the X-ray flux at their locations with exposures of at
most a few hundred seconds.  Table 1 lists the coordinates of the
three fields and the RASS field numbers obtained from the archive.

Figures 1, 2, and 3 show the RASS fields containing each target.  In
each field, circles surround the objects for which X-ray counts were
extracted.  The labelled circles correspond to the sources listed in
Table 3.  In Figure 2, two of the unlabelled, unextracted sources are
identified as Seyfert 1 galaxies (1RXS J221632.7-160533 
northernmost; 1RXS J221300.1-171019, westernmost source).

We extracted the counts at the positions of the target stars from the
data using apertures 5$'$ in radius.  The background was obtained from
an annulus of outer diameter 20$'$.  We also extracted the mean
exposure time from the corresponding location in the exposure map.
Table 2 lists the extracted counts.  Bayesian upper limits were
determined using the prescription in \cite{Lor92}.  

The estimated distances to two target stars fall in the range of 4-6
kpc \citep{Beers00}.  For simplicity, we adopt 5 kpc.  For BD+17, the
distance is listed as 0.7 kpc \citep{Beers00}.  The column density in
the direction of the targets is low, a few $\times$10$^{20}$ cm$^{-2}$
(Table 1).  Fluxes were computed from the 99\% upper limits assuming a
thermal spectrum of temperature 1 keV absorbed by the Galactic columns
in the direction of each source.  The fluxes approximately double if
the adopted temperature is 5 keV.

To provide additional confidence in our upper limits, the counts of
several faint point sources positionally near the uranium stars were
also extracted.  The positions and counts are listed in Table 3.  None
of the sources is an identified X-ray object based upon a search of
SIMBAD\footnote{This research has made use of the SIMBAD database,
operated by CDS, Strasbourg, France. SIMBAD access in the USA is
provided by funding from National Aeronautics and Space
Administration.} or NED\footnote{This research has made use of the
NASA/IPAC Extragalactic Database (NED) which is operated by the Jet
Propulsion Laboratory, California Institute of Technology, under
contract with the National Aeronautics and Space Administration.}.

\section{Discussion}

The {\it ROSAT} observations establish an upper limit at $\sim$1-1.5
keV of $\sim$10$^{32-33}$ ergs s$^{-1}$ for an assumed distance of 5 kpc.
This luminosity eliminates the presence of a black hole accreting at
the Eddington rate, assuming no very high local absorption, since the
expected flux from a 1 M$_{\odot}$ black hole would be easily detected
at 5 kpc, with an estimated flux rivaling that of the Crab.

If the binary did survive the explosion and if the compact object does
emit at typical Eddington values, a local high column provides an exit
from the restrictive upper limit.  Substantial local absorption, with
column densities $>$10$^{22}$ cm$^{-2}$, significantly remove X-rays
with energies below $\sim$2-2.5 keV.  As the {\it ROSAT} PSPC had
decreasing effective area above $\sim$2 keV, the sensitivity to hard
X-rays of the survey observations is considerably poorer than the flux
limit mentioned above.  Figure 4 illustrates the column
density-luminosity relation that yields no detection in the {\it
ROSAT} bandpass.

Other compact objects may exist in the assumed binaries, regardless of
whether we understand the overall evolutionary scenario.  In addition,
quiescent levels of some X-ray binaries in elliptical orbits (e.g.,
4U0115+63) are low with a typical L$_{\rm X}$ of
$\lesssim$2$\times$10$^{32}$ erg s$^{-1}$ \citep{Tam92}.  Persistent
X-ray binaries often have low L$_{\rm X}$, too.  For example, X Per
contains a neutron star, but has a total X-ray luminosity in the 2-10
keV band of 9${^{+8}_{-5}{\times}}$10$^{34}$ erg s$^{-1}$ \citep{S93}.

With relatively high upper limits, explanations for a non-detection
abound.  These include the failure of the assumption that the binary
survived the explosion, ``off'' states in wind- or Roche lobe-
accreting systems, or alternative explanations to generate the
r-process contaminants.  A larger discussion of these issues will be
presented with the expected improvement in the upper limits (see
following discussion).  We did search proper motion catalogs for each
of the targets.  Only BD+17 has a measured and published proper motion
of magnitude $\sim$52 milliarcsec yr$^{-1}$; the component motion in
right ascension differs between \cite{RB88} (-34 milliarcsec
yr$^{-1}$) and \cite{Beers00} (-46 milliarcsec yr$^{-1}$).  The proper
motions of the two CS stars have apparently not as yet been published.

A sensitive observation at energies above $\sim$3 keV would eliminate
much of all of the local column density argument.  A 10 ksec
observation with the XMM-Newton EPIC detector reaches L$_{\rm X}$
$\sim$2-5$\times$10$^{31}$ erg s$^{-1}$, constraining nearly all
compact binary scenarios including cataclysmic variables (magnetic CV
AM Her: L$_{\rm X}$ $\sim$2$\times$10$^{33}$ erg s$^{-1}$; dwarf nova
CV SS Cyg: L$_{\rm X}$ $\sim$1$\times$10$^{32}$ erg s$^{-1}$;
\cite{Warn95}) and neutron star wind accretors (e.g., 4U1700+24,
accreting from a red giant at ${\dot M}$ $\sim$10$^{-9}$ M$_{\odot}$
yr$^{-1}$, has a measured 2-10 keV luminosity of
$\sim$2$\times$10$^{32}$ erg $^{-1}$, \citealt{Mas01}).

The limit reaches the upper end of the luminosity function of RS CVn
stars \citep{SCS96,Demp93} but does not eliminate active binaries such
as BY Dra systems \citep{Demp97} which have L$_{\rm X}$
$\sim$10$^{28-30}$ erg s$^{-1}$.  The limit also does not reach single K
supergiants which have L$_{\rm X}$ $\sim$few $\times$10$^{29}$ erg
s$^{-1}$ (e.g., \citealt{Hun96}).

Two of the targets (CS31082 and CS22892) have been approved for Cycle 2
XMM observations, so more restrictive upper limits or detections will
be available shortly.

\acknowledgments

This research was supported by contract number NAS8-39073 to SAO.

\begin{figure}
\centering
\caption{Field of CS31082-001.  The circles are 5$'$ in radius and 
surround the star's location as well as ``test'' sources A and B.}
\label{fig1}
\scalebox{0.5}{\rotatebox{-90}{\includegraphics{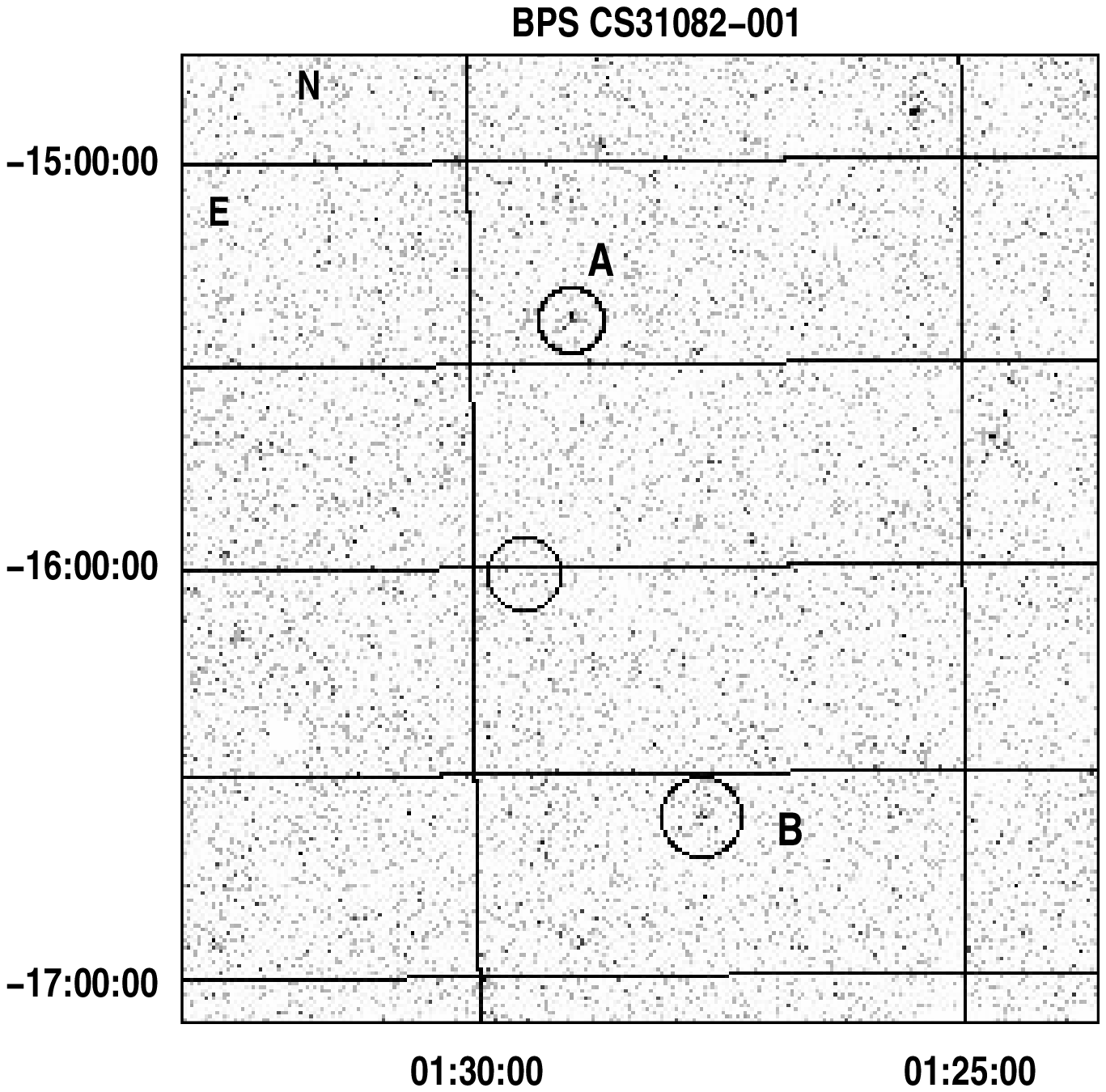}}}
\end{figure}

\begin{table*}
\small
\begin{center}
\caption{``Uranium'' Stars and RASS Fields}
\begin{tabular}{lccccrrrrr}
                 & \multicolumn{2}{c}{J2000} & ROSAT & N$_{\rm H}$ &  & Dist & Rad. & \multicolumn{2}{c}{Proper Motion\tablenotemark{b}} \\
  Star           &   RA       &    Dec    & Field\tablenotemark{a} & (cm$^{-2}$)  & [Fe/H] & (kpc) & Vel. & RA & Dec \\ \hline
BPS CS31082-001  & 01:29:31.2 & -16:00:48 & 932004 & 1.5(20) & -2.9 & $\cdots$ & $\cdots$ & $\cdots$ & $\cdots$ \\ 
BD +17$^{\circ}$3248 & 17:28:14.5 & +17:30:36 & 931446 & 5.9(20) & -2.0 & 0.703 & -146$\pm$1 & -46.7$\pm$0.8 & -24.2$\pm$0.7 \\
BPS CS22892-052  & 22:17:01.5 & -16:39:26 & 932059 & 2.6(20) & -3.1 & 5.7 & 10$\pm$10 & $\cdots$ & $\cdots$ \\ \hline
\end{tabular}
\end{center}
\tablenotetext{a}{identification number for {\it ROSAT} All-Sky
Survey field}
\tablenotetext{b}{Values for proper motion in milliarcsec yr$^{-1}$.}
\end{table*}

\begin{table*}
\begin{center}
\caption{Extracted X-ray Counts: ``Uranium'' Stars}
\begin{tabular}{lccrrrcc}
                 &       & 99\% & ExpTime &  & 99\% & \\
  Star           & Counts$\pm$Err & Counts & (sec) & Rate &Flux\tablenotemark{a} & L$_{\rm X}$\tablenotemark{b}\\ \hline
CS31082-001  & 2.2$\pm$2.4 & $<$6.9 & 435 & 0.016 & 1.8(-13) & 5.4(32) \\
BD+17$^{\circ}$ 3248  & 7.8$\pm$3.9 & $<$10.8 & 350 & 0.020 & 4.9(-13) & 9.3(31) \\
CS22892-052  & 5.8$\pm$3.0 & $<$11.2 & 277 & 0.042 & 6.0(-13) & 2.3(33) \\ \hline
\end{tabular}
\end{center}
\tablenotetext{a}{Counts converted to an unabsorbed flux assuming the
Galactic column toward the sources with a 1 keV bremsstrahlung
spectrum.  For a temperature of 5 keV, the fluxes are all $\sim$2
times higher.}
\tablenotetext{b}{L$_{\rm X}$ calculated assuming the distance in Table 1 or 5 kpc for CS31082.}
\end{table*}

\begin{table*}
\begin{center}
\caption{Extracted X-ray Counts for Test Sources}
\begin{tabular}{lrrcccr}
      & \multicolumn{2}{c}{J2000} &   Net  & ExpT &    &    \\
Field &   RA    &  Dec    & Counts$\pm$Err & (sec) & Rate & Flux\tablenotemark{a} \\ \hline
CS31082-A & 01:28:59.5 & -15:23:19 & 19.2$\pm$4.9 & 449 & 0.043$\pm$0.011 & 4.8$\pm$1.2(-13) \\
CS31082-B & 01:27:42.9 & -16:36:30 & 15.2$\pm$4.6 & 394 & 0.038$\pm$0.012 & 4.3$\pm$1.3(-13) \\ 
BD+17-A & 17:30:34.9 & +17:30:54 & 30.8$\pm$6.2 & 570 & 0.054$\pm$0.011 & 1.1$\pm$0.21(-12) \\
CS22892-A & 22:14:30.2 & -17:02:03 & 24.8$\pm$5.3 & 308 & 0.081$\pm$0.017 & 1.16$\pm$0.24(-12) \\ \hline
\end{tabular}
\end{center}
\tablenotetext{a}{Counts converted to an unabsorbed flux assuming the Galactic
column toward the sources with a 1 keV bremsstrahlung spectrum.}
\end{table*}

\newpage
\begin{figure}[t]
\centering
\caption{Field of CS22892-052.  The circles are 5$'$ in radius
and surround the star's location as well as ``test'' source A.  
The northern- and western-most uncircled sources are known Seyfert 1
galaxies; the southernmost is an unidentified RASS source (see text).}
\label{fig2}
\scalebox{0.5}{\rotatebox{-90}{\includegraphics{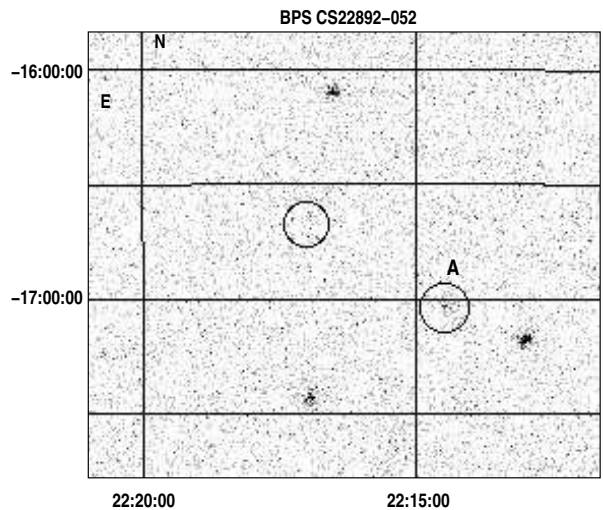}}}
\end{figure}

\newpage
\begin{figure}[t]
\centering
\caption{Field of BD+17$^{\circ}$ 3248.  The circles 
surround the star's location as well as test source 'A'.
The source in the lower right portion of the figure is an unidentified
RASS source (1RXS172534.0+160923).}
\label{fig3}
\scalebox{0.5}{\rotatebox{0}{\includegraphics{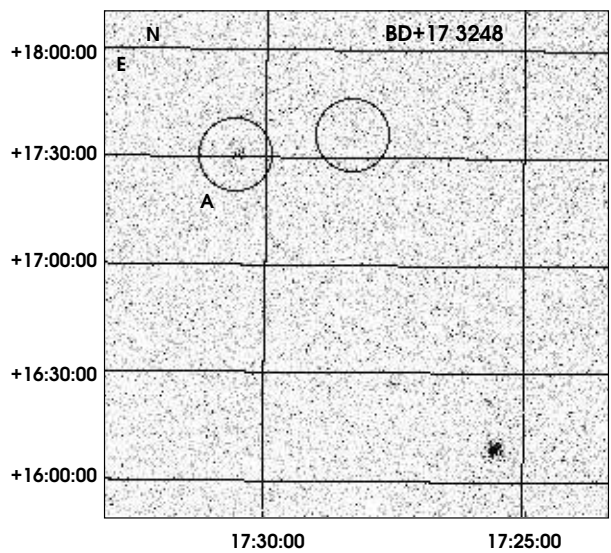}}}
\end{figure}

\newpage
\begin{figure*}[t]
\centering
\caption{The solid and dashed curves are the locus of column densities
that could absorb all the counts in the respective bands and
illustrate the difficulty of obscuring hard X-ray sources from the
hard band of XMM-Newton or Chandra (left vertical axis).  The dashed-dotted
curve shows the luminosities probed by a 25-count detection of a 1-keV
source assumed 5 kpc distant for exposure times on the right axis.}
\label{fig5}
\scalebox{0.5}{\rotatebox{0}{\includegraphics{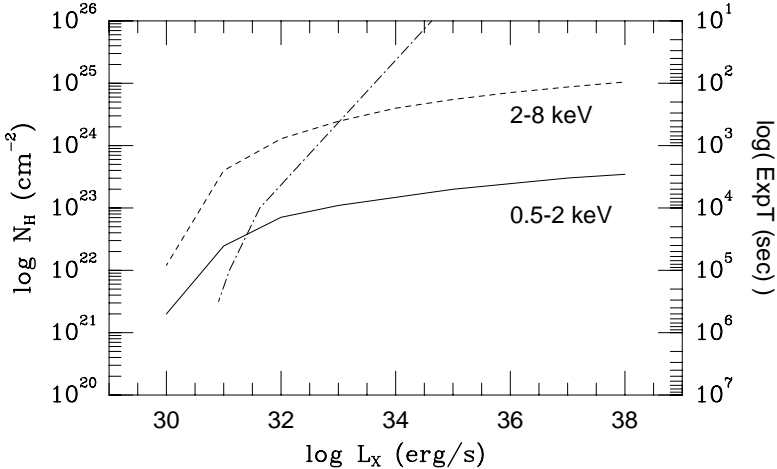}}}
\end{figure*}

\end{document}